\documentclass[aps,preprintnumbers,superscriptaddress,showpacs]{revtex4}
\usepackage{epsfig}
\usepackage{psfrag}
\usepackage{amsfonts}
\usepackage{graphicx}
\usepackage{dcolumn}
\usepackage{bm}
\usepackage[utf8]{inputenc}

\begin{document}

\title{Holographic Schwinger effect in spinning black hole backgrounds}

\author{Yi-ze Cai}\affiliation{School of Mathematics and Physics, China University
of Geosciences, Wuhan 430074, China}

\author{Zi-qiang Zhang}
\email{zhangzq@cug.edu.cn} \affiliation{School of Mathematics and
Physics, China University of Geosciences, Wuhan 430074, China}

\begin{abstract}
We perform the potential analysis for the holographic Schwinger
effect in spinning Myers-Perry black holes. We compute the
potential between the produced pair by evaluating the classical
action of a string attaching on a probe D3-brane sitting at an
intermediate position in the AdS bulk. It turns out that
increasing the angular momentum reduces the potential barrier thus
enhancing the Schwinger effect, consistent with previous findings
obtained from the local Lorentz transformation. In particular,
these effects are more visible for the particle pair lying in the
transversal plane compared with that along the longitudinal
orientation. In addition, we discuss how the Schwinger effect
changes with the shear viscosity to entropy density ratio at
strong coupling under the influence of angular momentum.
\end{abstract}
\pacs{11.25.Tq, 11.15.Tk, 11.25-w}

\maketitle
\section{Introduction}
In the vacuum of quantum electrodynamics (QED), virtual
electron-position pairs could be materialized and become real
particles in the presence of a strong electric-field. This
phenomenon has been termed Schwinger effect. The production rate
$\Gamma$ of electron-positron pairs for the case of weak-coupling
and weak-field was first studied by Schwinger in 1951 \cite{JS}
\begin{equation}
\Gamma\sim exp\Big({\frac{-\pi m^2}{eE}}\Big),\label{gama0}
\end{equation}
where $m$, $e$, $E$ are an electron mass, an elementary electric
charge and an external electric field, respectively. One can see
that there is no critical field in (\ref{gama0}). Later, the
calculation of $\Gamma$ for the case of arbitrary-coupling and
weak-field was considered in \cite{IK}
\begin{equation}
\Gamma\sim exp\Big({\frac{-\pi
m^2}{eE}+\frac{e^2}{4}}\Big),\label{gama00}
\end{equation}
one can find that there is a critical electric field at
$eE_c=(4\pi/e^2)m^2\simeq137m^2$ in (\ref{gama00}), but this value
doesn't meet the weak-field condition $eE\ll m^2$. Therefore, it
seems that one could hardly get the critical field under
weak-field condition.

In fact, Schwinger effect is not limited to QED but ubiquitous for
quantum field theory (QFT) coupled to an U(1) gauge field.
However, it is difficult to deal with this issue with the standard
method in QFT. A possible way is to employ the AdS/CFT
correspondence
\cite{Maldacena:1997re,Gubser:1998bc,MadalcenaReview}. Using
AdS/CFT, Semenoff and Zarembo pioneered the holographic Schwinger
effect in 2011 \cite{GW}. They pointed out that a supersymmetric
Yang-Mills (SYM) coupled with an U(1) gauge field can be realized
by breaking the gauge group from $SU(N+1)$ to $SU(N)\times U(1)$
via the Higgs mechanism. In doing so, Schwinger effect could be
modelled in the higgsed $\mathcal N=4$ SYM. The production rate of
the fundamental particles (W-boson supermultiplet or quarks) at
large $N_c$ (color number) and large $\lambda$ ('t Hooft coupling)
are evaluated as \cite{GW}
\begin{equation}
\Gamma\sim
exp\Big[-\frac{\sqrt{\lambda}}{2}\Big(\sqrt{\frac{E_c}{E}}-\sqrt{\frac{E}{E_c}}\Big)^2\Big],
\qquad E_c=\frac{2\pi m^2}{\sqrt{\lambda}},
\end{equation}
interestingly, the value of $E_c$ coincides with the DBI result
\cite{GW}. Subsequently, there are many works to develop and
extend this idea. For instance, the universal aspects of
holographic Schwinger effect for general backgrounds was discussed
in \cite{YS2}. The holographic Schwinger effect with constant
electric and magnetic fields was studied in \cite{SB,YS3}.
Furthermore, the potential barrier for holographic Schwinger
effect was explored in various backgrounds
\cite{YS,YS1,KB,MG,ZQ,ZQ1,LS,ZR,swe}. Other related results can be
found in \cite{DK,SCH,KHA,KHA1,KG,LS1,yz,yd,swe1,WF}.

In this work, we shall explore the holographic Schwinger effect in
spinning black hole backgrounds by means of the AdS/CFT
correspondence. The motivations are as follows: First, Schwinger
effect may have a connection with the heavy-ion collisions
experiments, where strong electro-magnetic fields and color fields
could be induced due to the collision of heavy ions. Second, it
has been reported that the noncentral collisions tends to deposit
high angular momentum in the quark gluon plasma (QGP) generated in
heavy ion collisions and such angular momentum may give rise to
significant observable effects in QGP \cite{nat,zt,fb,xg,lg}.
Already, various observables with respect to QGP have been studied
under the influence of angular momentum from holography. Such as
jet quenching parameter \cite{js,bmc}, drag force
\cite{iy,iy1,ana}, energy loss \cite{kb,mat,df},
confinement/deconfinement \cite{nr,xc} and running coupling
constant \cite{xc1}. Other related results can be found in
\cite{sb,bm,bm1,hb,an,sy,aa}. In previous literature, we have
investigated the holographic Schwinger effect in a soft wall model
\cite{yz} by taking a local Lorentz transformation \cite{mb,ce,am}
to the static frame of a small segment of the rotating medium. But
this approach has its limitations: its metric can only describe a
small neighbourhood around $l=l_0$ and a domain less than $2\pi$
wide of the rotating medium \cite{jx}, where $l_0$ represents the
radius to the rotating axis. Are there any other ways besides the
local Lorentz transformation to mimic the rotating QGP? One
possible approach is to utilize the rotating black holes, e.g.,
Kerr-AdS$_5$ metric \cite{sw}. Recently, the shear viscosity to
entropy density ratio $\eta/s$ has been calculated in
five-dimensional Myers-Perry black holes. These black holes are a
form of spinning five-dimensional AdS black holes and have been
found as vacuum solutions within Einstein gravity \cite{mg}. In
particular, for these solutions, the boundary is compact and the
dual SYM lives on $S^3\times \mathbb{R}$. So if one is interested
in a dual to a spinning fluid, e.g., QGP, in flat space
$\mathbb{R}^{3,1}$, one could consider the large black holes.
Furthermore, if one tends to think of a regime of large
temperature in order to have a dual field theory on a non-compact
spacetime (this case maybe more relevant for applications to heavy
ion collisions), one could consider the planar limit black brane
as a limit of the large black holes. For these reasons, we would
like to reexamine the holographic Schwinger effect in
five-dimensional Myers-Perry black holes. We want to see whether
the results obtained from these spinning black hole backgrounds
are in line with those from the local Lorentz transformation.
Also, by comparing with the results of \cite{mg}, we want to see
how the Schwinger effect changes with $\eta/s$ at strong coupling
under the influence of angular momentum.

The organization of the paper is as follows. In the next section,
we briefly recall the spinning Myers-Perry black holes given in
\cite{sw,gw,gw1}. In section 3, we perform the potential analysis
for the holographic Schwinger effect in these backgrounds and
analyze how angular momentum modifies the production rate.
Finally, we give our conclusions and discussions in section 4.

\section{Setup}
The more familiar metric of the five-dimensional spinning black
holes is written by Hawking et al. \cite{sw}
\begin{eqnarray}
ds^{2}&=&-\frac{\Delta }{\rho ^{2}}(dt_{H}-\frac{a\sin^{2}\theta
_{H}}{\Xi _{a}}d\phi _{H}-\frac{b\cos^{2}\theta _{H}}{\Xi
_{b}}d\psi _{H})^{2}\nonumber\\&+&\frac{\Delta _{\theta
_{H}}\sin^{2}\theta _{H}}{\rho
^{2}}(adt_{H}-\frac{r_{H}^{2}+a^{2}}{\Xi _{a}}d\phi
_{H})^{2}+\frac{\Delta _{\theta _{H}}\cos^{2}\theta _{H}}{\rho
^{2}}(bdt_{H}-\frac{r_{H}^{2}+b^{2}}{\Xi _{b}}d\psi
_{H})^{2}+\frac{\rho ^{2}}{\Delta
}dr_{H}^{2}\nonumber\\&-&\frac{\rho ^{2}}{\Delta _{\theta
_{H}}}d\theta
_{H}^{2}+\frac{1+\frac{r_{H}^{2}}{L^{2}}}{r_{H}^{2}\rho
^{2}}(abdt_{H}-\frac{b(r^{2}+a^{2})\sin^{2}\theta _{H} }{\Xi
_{a}}d\phi _{H}-\frac{a(r^{2}+b^{2})\cos^{2}\theta _{H}}{\Xi
_{b}}d\psi _{H})^{2},\label{metric1}
\end{eqnarray}
with
\begin{eqnarray}
\Delta &=& \frac{1}{r_{H}^{2}}(r_{H}^{2}+a^{2})(r_{H}^{2}+b^{2})(1+\frac{r_{H}^{2}}{L^{2}})-2M,\nonumber\\
\Delta _{\theta _{H}} &=& 1-\frac{a^{2}}{L^{2}}\cos^{2}\theta _{H}-\frac{b^{2}}{L^{2}}\sin^{2}\theta _{H},\nonumber\\
\rho &=&r_{H}^{2}+a^{2}\cos^{2}\theta _{H}+b^{2}\sin^{2}\theta _{H},\nonumber\\
\Xi _{a}&=&1-\frac{a^{2}}{L^{2}},\nonumber\\
\Xi _{b}&=&1-\frac{b^{2}}{L^{2}},
\end{eqnarray}
where $t_{H}$ is the time, $L$ is the AdS radius, $r_{H}$
represents the AdS radial coordinate, $(\phi _{H},\psi _{H},\theta
_{H})$ are the angular Hopf coordinates. ${a,b}$ denote two
independent angular momentum parameters which could generate all
possible rotations. Here we will focus on the case of $a=b$, which
is referred as the simply spinning Myers-Perry black holes
\cite{gw,gw1}.

In order to analyze conveniently, one can employ more convenient
coordinates and reparameterize the mass following \cite{km}
\begin{eqnarray}
t&=&t_{H},\nonumber\\
r^{2}&=& \frac{a^{2}+r_{H}^{2}}{1-\frac{a^{2}}{L^{2}}},\nonumber\\
\theta &=& 2\theta _{H},\nonumber\\
\phi &=& \phi _{H}-\psi _{H},\nonumber\\
\psi &=& -\frac{2at_{H}}{L^{2}}+\phi _{H}+\psi _{H},\nonumber\\
b &=& a,\nonumber\\
\mu &=& \frac{M}{(L^{2}-a^{2})^{3}}.\nonumber\\
\end{eqnarray}

Then the metric (\ref{metric1}) can be simplified to
\begin{equation}
ds^{2}=-(1+\frac{r^{2}}{L^{2}})dt^{2}+\frac{dr^{2}}{G(r)}+\frac{r^{2}}{4}((\sigma
^{1})^{2}+(\sigma ^{2})^{2}+(\sigma ^{3})^{2})+\frac{2\mu
}{r^{2}}(dt+\frac{a}{2}\sigma ^{3})^{2},\label{metric2}
\end{equation}
with
\begin{eqnarray}
G(r)&=&1+\frac{r^{2}}{L^{2}}-\frac{2\mu (1-\frac{a^{2}}{L^{2}})}{r^{2}}+\frac{2\mu a^{2}}{r^{4}},\nonumber\\
\mu &=&\frac{r_{h}^{4}(L^{2}+r_{h}^{2})}{2L^{2}r_{h}^{2}-2a^{2}(L^{2}+r_{h}^{2})},\nonumber\\
\sigma ^{1} &=& -\sin\psi d\theta +\cos\psi \sin\theta d\phi ,\nonumber\\
\sigma ^{2} &=& \cos\psi d\theta +\sin\psi \sin\theta d\phi ,\nonumber\\
\sigma ^{3} &=& d\psi +\cos\theta d\phi.
\end{eqnarray}
where the range of the coordinates is
\begin{equation}
-\infty <t< \infty ,\ r_{h}< r< \infty ,\ 0\leq \theta \leq \pi ,\
0\leq \phi\leq 2\pi ,\ 0\leq  \psi < 4\pi .
\end{equation}
here $r_h$ is the outer horizon, defined by $G(r_h)=0$. It should
be noted that the Myers-Perry black holes defined by
(\ref{metric2}) have two instabilities \cite{km}. First, a
superradient instability has been found, which occurs at large
angular velocities $|\Omega L|>1$. In order to avoid this, we
consider $|\Omega L|<1$ here. The second instability
(Gregory-Laflamme instability) was found at small horizon radius
$r_h\sim L$. This instability is not within the range of
parameters which we consider $r_h\gg L$.

As prophesied above, in this work we are most interested in
rotating QGP, so we tend to consider the large black hole limit,
since this limit would be more relevant for applications to heavy
ion collisions \cite{mg}. For this purpose, one adopts the
following coordinate transformation
\begin{eqnarray}
t&=&\tau ,\nonumber\\
\frac{L}{2}(\phi -\pi )&=&x,\nonumber\\
\frac{L}{2}\tan(\theta  -\frac{\pi }{2} )&=&y,\nonumber\\
\frac{L}{2}(\psi -2\pi )&=&z,\nonumber\\
r&=&\widetilde{r},
\end{eqnarray}
then the coordinates in the new $(\tau ,\widetilde{r},x,y,z)$
coordinates become
\begin{eqnarray}
\tau  &\rightarrow& \beta ^{-1}\tau ,\nonumber\\
x &\rightarrow& \beta ^{-1}x,\nonumber\\
y &\rightarrow& \beta ^{-1}y,\nonumber\\
z &\rightarrow& \beta ^{-1}z,\nonumber\\
\widetilde{r} &\rightarrow& \beta \widetilde{r},\nonumber\\
\widetilde{r_{h}} &\rightarrow& \beta  \widetilde{r_{h}},\ (\beta
\rightarrow \infty )
\end{eqnarray}
where $\beta $ is an appropriate power of a scaling factor.

As a result, one obtains a Schwarzschild black brane metric that
has been boosted about the $\tau$-$z$ plane
\begin{equation}
ds^{2}=\frac{r^{2}}{L^{2}}(-d\tau
^{2}+dx^{2}+dy^{2}+dz^{2}+\frac{r_{h}^{4}}{r^{4}(1-\frac{a^{2}}{L^{2}})}(d\tau
+\frac{a}{L}dz)^{2})+\frac{L^{2}r^{2}}{r^{4}-r_{h}^{4}}dr^{2},\label{metric3}
\end{equation}
note that for $a=0$ in (\ref{metric3}), the Schwarzschild black
brane is reproduced.

The temperature of this boosted black brane reads
\begin{equation}
T=\frac{r_{h}\sqrt{L^{2}-a^{2}}}{\pi L^{3}}.
\end{equation}

Incidentally, the $\eta/s$ in spinning Myers-Perry black holes is
given by \cite{mg}
\begin{eqnarray}
\frac{\eta_\perp}{s}&=&\frac{1}{4\pi},\nonumber\\
\frac{\eta_\parallel}{s}&=&\frac{1}{4\pi}(1-a^2),\label{eta}
\end{eqnarray}
one can see that $\eta/s$ depends on the angle between the spatial
direction of the measurement and the angular momentum. For more
details about the spinning Myers-Perry black holes, we refer to
\cite{sw,gw,gw1,km}.

\section{Potential analysis in holographic Schwinger effect}
In this section we investigate the behavior of the Schwinger
effect for the background (\ref{metric3}) following \cite{YS}. The
Nambu-Goto action is
\begin{equation}
S=T_F\int d\xi d\eta\mathcal L=T_F\int d\xi d\eta \sqrt{g},
\label{S}
\end{equation}
where $T_F=\frac{1}{2\pi\alpha^\prime}$ is the fundamental string
tension. $\alpha^\prime$ is related to $\lambda$ via
$\frac{L^2}{\alpha^\prime}=\sqrt{\lambda}$. $g$ denotes the
determinant of the induced metric
\begin{equation}
g_{\alpha\beta}=g_{\mu\nu}\frac{\partial
X^\mu}{\partial\sigma^\alpha} \frac{\partial
X^\nu}{\partial\sigma^\beta},
\end{equation}
with $g_{\mu\nu}$ and $X^\mu$ being the metric and target space
coordinate, respectively.

It can be seen from (\ref{metric3}) that the boost exists in the
$\tau$-$z$ plane, implying the angular momentum can distinguish
the different orientations of the particle pair e.g.,$(Q\bar{Q})$
axis with respect to the direction of rotation (defined here to be
$z$ axis). Two extreme cases are worthy of note: transverse case
(the pair's axis is on the $x-y$ plane) and parallel case (the
pair's axis is on the $z$ axis). Next, we will examine the two
cases in turn.

\subsection{Transverse to rotation direction}
First we consider the transverse case. Without loss of generality,
one could assume that the pair's axis is along the $x$ direction,
\begin{equation}
\tau=\xi, \qquad x=\eta, \qquad y=0, \qquad z=0, \qquad
r=r(\eta).\label{t,x}
\end{equation}

Given that, the induced metric can be written as
\begin{equation}
g_{00}=-\frac{r^{2}}{L^{2}}+\frac{r_{h}^{4}}{r^{2}(L^{2}-a^{2})},\qquad
g_{01}=g_{10}=0,\qquad
g_{11}=\frac{r^{2}}{L^{2}}+\frac{L^{2}r^{2}}{r^{4}-r_{h}^{4}}\dot{r}^{2},
\end{equation}
where $\dot{r}=\frac{dr}{d\eta}$.

The Lagrangian density reads
\begin{equation}
\mathcal{L}=\sqrt{A(r)+B(r)\dot{r}^{2}}, \label{L}
\end{equation}
with
\begin{equation}
A(r)=\frac{r^{4}}{L^{4}}-\frac{r_{h}^{4}}{L^{2}(L^{2}-a^{2})},\qquad
B(r)=-\frac{L^{2}r_{h}^{4}}{(L^{2}-a^{2})(r^{4}-r_{h}^{4})}+\frac{r^{4}}{r^{4}-r_{h}^{4}}.
\end{equation}

One can see that $\mathcal L$ does not depend on $\eta$
explicitly, so the Hamiltonian is conserved,
\begin{equation}
\mathcal L-\frac{\partial\mathcal
L}{\partial\dot{r}}\dot{r}=Constant.
\end{equation}

Imposing the boundary condition at $\eta=0$
\begin{equation}
\frac{dr}{d\eta}=0,\qquad  r=r_c\qquad (r_h<r_c)\label{con},
\end{equation}
one gets
\begin{equation}
\frac{dr}{d\eta}=\sqrt{\frac{A^2(r)-A(r)A(r_c)}{A(r_c)B(r)}}\label{dotr},
\end{equation}
where $A(r_c)=A(r)|_{r=r_c}$.

Integrating (\ref{dotr}), the inter-distance between the particle
pair is obtained
\begin{equation}
x^\perp=2\int_{r_c}^{r_0}dr\sqrt{\frac{A(r_c)B(r)}{A^2(r)-A(r)A(r_c)}}\label{xx},
\end{equation}
where we have placed the probe D3-brane at an intermediate
position $r=r_0$ rather than close to the boundary. Such
operations could yield a finite mass which then makes sense of the
production rate \cite{GW}.

Substituting (\ref{L}), (\ref{dotr}) into (\ref{S}), the sum of
Coulomb potential and static energy is obtained
\begin{equation}
V_{CP+E}=2T_F\int_{r_c}^{r_0}dr\sqrt{\frac{A(r)B(r)}{A(r)-A(r_c)}}.\label{en}
\end{equation}

To proceed, we calculate the critical field. The DBI action is
\begin{equation}
S_{DBI}=-T_{D3}\int
d^4x\sqrt{-\det(G_{\mu\nu}+\mathcal{F}_{\mu\nu})}\label{dbi},
\end{equation}
with
\begin{equation}
T_{D3}=\frac{1}{g_s(2\pi)^3\alpha^{\prime^2}}, \qquad
\mathcal{F}_{\mu\nu}=2\pi\alpha^\prime F_{\mu\nu},
\end{equation}
where $T_{D3}$ refers to the D3-brane tension.

Assuming the electric field is turned on along the $x$ direction
\cite{YS}, one gets
\begin{equation}
G_{\mu\nu}+\mathcal{F}_{\mu\nu}=\left(
\begin{array}{cccc}
-\frac{r^{2}}{L^{2}}+\frac{r_{h}^{4}}{r^{2}(L^{2}-a^{2})} & 2\pi\alpha^\prime E & 0 &\frac{ar_{h}^{4}}{r^{2}L(L^{2}-a^{2})}\\
 -2\pi\alpha^\prime E & \frac{r^{2}}{L^{2}} & 0 & 0 \\
 0 & 0 & \frac{r^{2}}{L^{2}} & 0\\
\frac{ar_{h}^{4}}{r^{2}L(L^{2}-a^{2})} & 0 & 0 &
\frac{r^{2}}{L^{2}}+\frac{r_{h}^{4}a^{2}}{r^{2}L^{2}(L^{2}-a^{2})}
\end{array}
\right),
\end{equation}
yielding
\begin{equation}
\det(G_{\mu\nu}+\mathcal{F}_{\mu\nu})=\frac{r^{2}}{L^{2}}[(2\pi
{\alpha
}'E)^{2}(\frac{r^{2}}{L^{2}}+\frac{r_{h}^{4}a^{2}}{L^{2}r^{2}(L^{2}-a^{2})})+\frac{r^{2}r_{h}^{4}}{L^{4}(L^{2}-a^{2})}
-\frac{r^{2}r_{h}^{4}a^{2}}{L^{6}(L^{2}-a^{2})}-\frac{r^{6}}{L^{6}}].\label{det}
\end{equation}

Plugging (\ref{det}) into (\ref{dbi}) and making the probe
D3-brane located at $r=r_0$, one gets
\begin{equation}
S_{DBI}=-T_{D3}\frac{r_{0}}{L}\int d^4x
\sqrt{\frac{r_{0}^{2}r_{h}^{4}a^{2}}{L^{6}(L^{2}-a^{2})}+\frac{r_{0}^{6}}{L^{6}}-\frac{r_{0}^{2}r_{h}^{4}}{L^{4}(L^{2}-a^{2})}-(2\pi
{\alpha
}'E)^{2}(\frac{r_{0}^{2}}{L^{2}}+\frac{r_{h}^{4}a^{2}}{L^{2}r_{0}^{2}(L^{2}-a^{2})})}\label{dbi1}.
\end{equation}

To avoid the action (\ref{dbi1}) being ill-defined, one needs
\begin{equation}
\frac{r_{0}^{2}r_{h}^{4}a^{2}}{L^{6}(L^{2}-a^{2})}+\frac{r_{0}^{6}}{L^{6}}-\frac{r_{0}^{2}r_{h}^{4}}{L^{4}(L^{2}-a^{2})}-(2\pi
{\alpha
}'E)^{2}(\frac{r_{0}^{2}}{L^{2}}+\frac{r_{h}^{4}a^{2}}{L^{2}r_{0}^{2}(L^{2}-a^{2})})\geq0,\label{ec}
\end{equation}
which leads to
\begin{equation}
E\leq
T_{F}\sqrt{\frac{\frac{r_{0}^{6}}{L^{4}}+\frac{r_{0}^{2}r_{h}^{4}a^{2}}{L^{4}(L^{2}-a^{2})}
-\frac{r_{0}^{2}r_{h}^{4}}{L^{2}(L^{2}-a^{2})}}{r_{0}^{2}+\frac{r_{h}^{4}a^{2}}{r_{0}^{2}(L^{2}-a^{2})}}}.
\end{equation}

As a result, the critical field is
\begin{equation}
E_c^\perp=T_{F}\sqrt{\frac{\frac{r_{0}^{6}}{L^{4}}+\frac{r_{0}^{2}r_{h}^{4}a^{2}}{L^{4}(L^{2}-a^{2})}-\frac{r_{0}^{2}r_{h}^{4}}{L^{2}(L^{2}-a^{2})}}{r_{0}^{2}
+\frac{r_{h}^{4}a^{2}}{r_{0}^{2}(L^{2}-a^{2})}}},\label{ec1}
\end{equation}
one can see that $E_c^\perp$ depends on $T$, $r_0$ and $a$.

Finally, the total potential for the transverse case can be
written as
\begin{eqnarray}
V_{tot}^\perp(x)&=&V_{CP+E}-E
x^\perp\nonumber\\&=&2pr_0T_F\int_1^{1/p}dy\sqrt{\frac{A(y)B(y)}{A(y)-A(y_c)}}\nonumber\\&-&
2pr_0T_F\alpha\sqrt{\frac{\frac{r_{0}^{6}}{L^{4}}+\frac{r_{0}^{2}(qr_{0})^{4}a^{2}}{L^{4}(L^{2}-a^{2})}-\frac{r_{0}^{2}(qr_{0})^{4}}{L^{2}(L^{2}
-a^{2})}}{r_{0}^{2}+\frac{(qr_{0})^{4}a^{2}}{r_{0}^{2}(L^{2}-a^{2})}}}\int_1^{1/p}dy\sqrt{\frac{A(y_c)B(y)}{A^2(y)-A(y)A(y_c)}},
\label{V}
\end{eqnarray}
where
\begin{equation}
\alpha\equiv\frac{E}{E_c^\perp}, \qquad
y\equiv\frac{r}{r_c},\qquad p\equiv\frac{r_c}{r_0},\qquad
q\equiv\frac{r_h}{r_0}, \label{afa}
\end{equation}
\begin{eqnarray}
A(y)&=&\frac{(pr_{0}y)^{4}}{L^{4}}-\frac{(qr_{0})^{4}}{L^{2}(L^{2}-a^{2})},\nonumber\\
B(y)&=&-\frac{L^{2}(qr_{0})^{4}}{(L^{2}-a^{2})((pr_{0}y)^{4}-(qr_{0})^{4})}+\frac{(pr_{0}y)^{4}}{(pr_{0}y)^{4}-(qr_{0})^{4}},\nonumber\\
A(y_c)&=&\frac{(pr_{0})^{4}}{L^{4}}-\frac{(qr_{0})^{4}}{L^{2}(L^{2}-a^{2})}.
\end{eqnarray}

The analysis of (\ref{V}) will be provided together with the
parallel case later.

\subsection{Parallel to rotation direction}
Now let's move on to the parallel case. Assuming the the particle
pair's axis is aligned in the $z$ direction,
\begin{equation}
\tau=\xi, \qquad x=0, \qquad y=0,\qquad z=\eta,\qquad
r=r(\eta).\label{t,z}
\end{equation}

Through similar calculations, the inter-distance, the critical
electric field and the total potential are obtained as
\begin{equation}
x^\parallel=2pr_{0}\int_{1}^{\frac{1}{p}}dy\sqrt{\frac{A_1(y_c)B_1(y)}{A_1^2(y)-A_1(y)A_1(y_c)}},\label{xx2}
\end{equation}
\begin{equation}
E_{c}^\parallel=T_{F}\sqrt{\frac{r_{0}^{4}}{L^{4}}+\frac{(qr_{0})^{4}a^{2}}{L^{4}(L^{2}-a^{2})}-\frac{(qr_{0})^{4}}{L^{2}(L^{2}-a^{2})}},\label{ec2}
\end{equation}
\begin{eqnarray}
V_{tot}^\parallel(x)&=&2pr_0T_F\int_1^{1/p}dy\sqrt{\frac{A_1(y)B_1(y)}{A_1(y)-A_1(y_c)}}\nonumber\\&-&
2pr_0T_F\alpha\sqrt{\frac{r_{0}^{4}}{L^{4}}+\frac{(qr_{0})^{4}a^{2}}{L^{4}(L^{2}-a^{2})}-\frac{(qr_{0})^{4}}{L^{2}(L^{2}-a^{2})}}\int_1^{1/p}dy
\sqrt{\frac{A_1(y_c)B_1(y)}{A_1^2(y)-A_1(y)A_1(y_c)}}, \label{V2}
\end{eqnarray}
with
\begin{eqnarray}
A_1(y)&=&\frac{(pr_{0}y)^{4}}{L^{4}}-\frac{(qr_{0})^{4}}{L^{2}(L^{2}-a^{2})}+\frac{a^{2}(qr_{0})^{4}}{L^{4}(L^{2}-a^{2})},\nonumber\\
B_1(y)&=&\frac{(pr_{0}y)^{4}}{(pr_{0}y)^{4}-(qr_{0})^{4}}-\frac{L^{2}(qr_{0})^{4}}{(L^{2}-a^{2})((pr_{0}y)^{4}-(qr_{0})^{4})},\nonumber\\
A_1(y_{c})&=&\frac{(pr_{0})^{4}}{L^{4}}-\frac{(qr_{0})^{4}}{L^{2}(L^{2}-a^{2})}+\frac{a^{2}(qr_{0})^{4}}{L^{4}(L^{2}-a^{2})},\nonumber\\
\label{V1}
\end{eqnarray}
where $y,p,q$ are the same as in (\ref{afa}) and $\alpha\equiv
E/E_c^\parallel$. We have checked that by plugging $a=0$ in
(\ref{V}) or (\ref{V1}), the results of SYM (without rotation)
\cite{YS} can be reproduced.

\begin{figure}
\centering
\includegraphics[width=8cm]{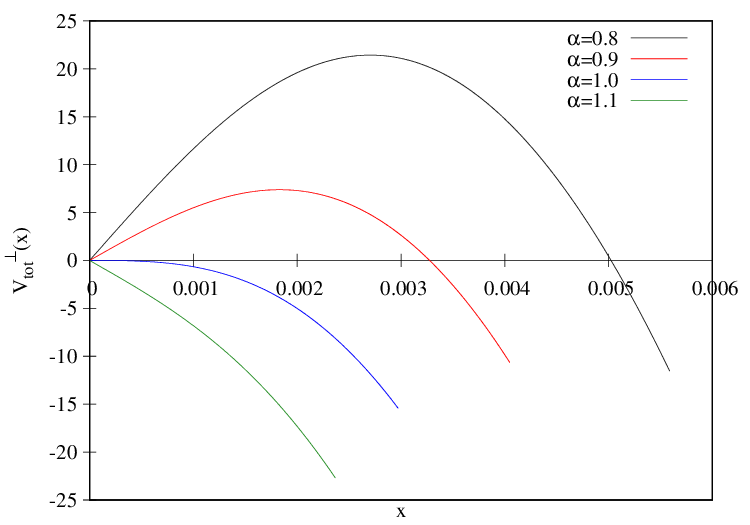}
\includegraphics[width=8cm]{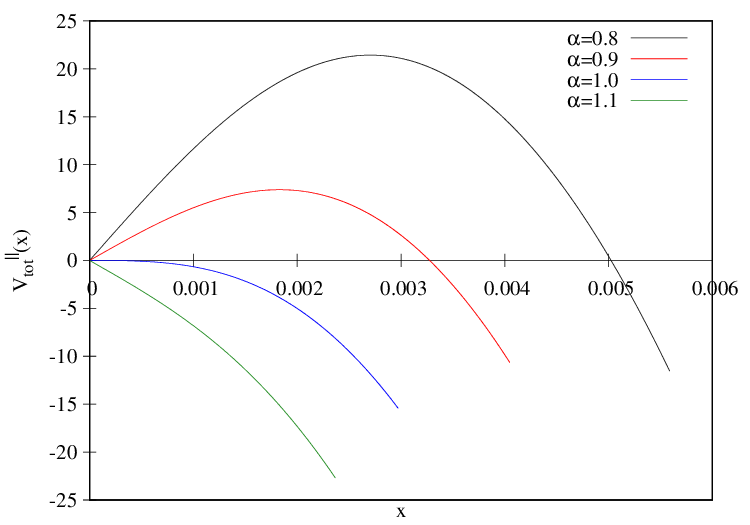}
\caption{$V_{tot}(x)$ versus $x$ with $a=0.2$. Left: Transverse
case. Right: Parallel case. In both panels from top to bottom
$\alpha=0.8, 0.9, 1, 1.1$, respectively.}
\end{figure}

\begin{figure}
\centering
\includegraphics[width=8cm]{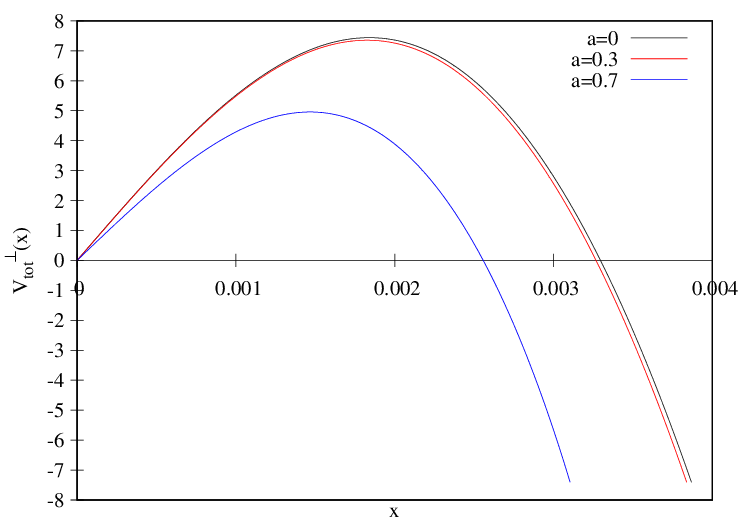}
\includegraphics[width=8cm]{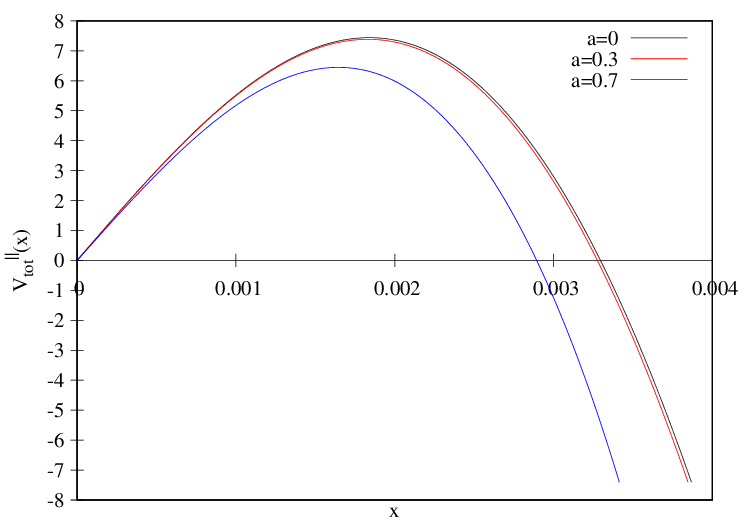}
\caption{$V_{tot}(x)$ versus $x$ with $\alpha=0.9$ for different
values of $a$. Left: Transverse case. Right: Parallel case. In
both panels from top to bottom $a=0, 0.3, 0.7$, respectively.}
\end{figure}

\begin{figure}
\centering
\includegraphics[width=8cm]{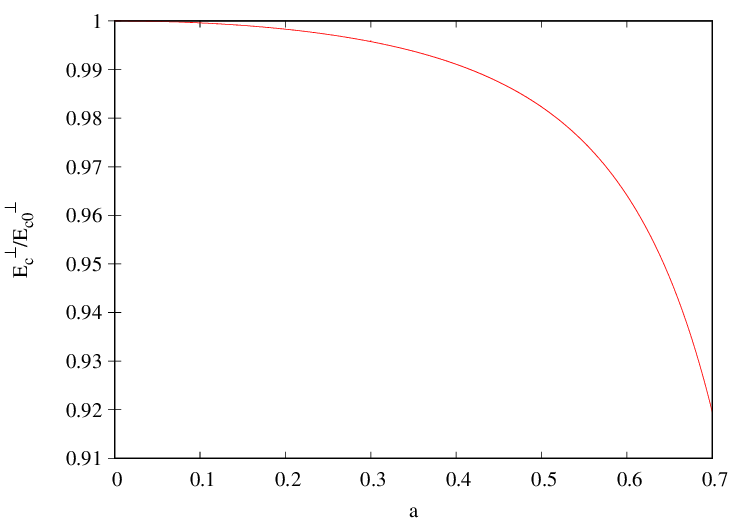}
\includegraphics[width=8cm]{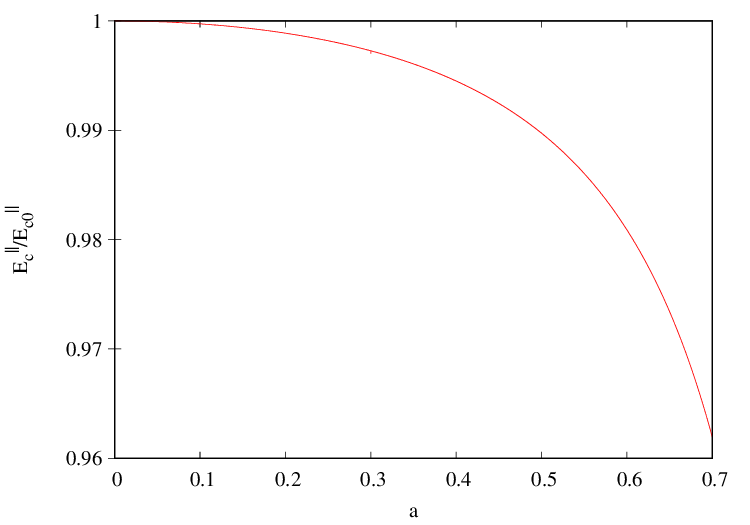}
\caption{$E_c/E_{c0}$ versus $a$. Left: Transverse case. Right:
Parallel case.}
\end{figure}

Before going on, we determine the values of some parameters.
First, we take $T_F=L=1$, similar to \cite{YS}. Moreover, we
choose a large (fixed) temperature of $T=100/\pi$, performing a
planar limit on the geometry yielding a black brane, as follows
from \cite{mg}. In addition, it has been suggested \cite{mg} that
the spinning black brane (\ref{metric3}) would be unstable at
sufficiently large angular momentum $a\approx 0.75L$. To alleviate
this, we consider $a<0.75L$ in calculations.

Let's discuss results. In fig.1, we plot $V_{tot}(x)$ as a
function of $x$ for different values of $\alpha$ with fixed
$a=0.2$ (other cases with different $a$ have similar picture),
where the left panel is for the transverse case while the right is
for the parallel case. From both panels, one can see that for
$\alpha<1$ (or $E<E_c$), the potential barrier is present and the
Schwinger effect can occur as tunneling process. With the increase
of $E$, the potential barrier decreases and finally vanishes at
$\alpha=1$ (or $E=E_c$). For $\alpha>1$ ($E>E_c$), the vacuum
becomes unstable catastrophically. These results fall in line with
\cite{YS}.

To understand how angular momentum modifies the Schwinger effect,
we plot $V_{tot}(x)$ against $x$ for different values of $a$ with
fixed $\alpha=0.9$ in fig.2, where the left panel is for the
transverse case while the right is for the parallel case. In both
panels from top to bottom $a=0, 0.3, 0.7$, respectively. From
these figures it is clear to see that as $a$ increases the height
and width of the potential barrier both decrease. As we know, the
higher (or the wider) the potential barrier, the harder the
produced pairs escape to infinity. One can thus conclude that the
inclusion of angular momentum decreases the potential barrier thus
enhancing the Schwinger effect. In other words, the presence of
angular momentum enhances the production rate. These results are
consistent with previous findings obtained from a soft wall model
\cite{yz}. Moreover, by comparing the two panels, one finds
angular momentum has important effect for the transverse case
comparing with the parallel case.

Also, one can examine how angular momentum affects the critical
electric field. To this end, we plot $E_c/E_{c0}$ versus $a$ in
fig.3, where the left panel is for the transverse case while the
right is for the parallel case, and $E_{c0}$ represents the
critical electric field at $a=0$. One can see that $E_c/E_{c0}$
decreases as $a$ increases. In particular, when $a=0.7$, the ratio
decreased by about 8 percent for the transverse case and 4 percent
for the parallel case. It is known that the smaller the critical
electric field, the easier the tunneling process. This is in
agreement with the previous potential analysis.

\section{Conclusion and discussion}
In this paper, we investigated the effect of angular momentum on
holographic Schwinger effect in spinning Myers-Perry black holes.
Along with the prescription in \cite{YS}, we calculated the
potential between the produced pair by evaluating the classical
action of a string attaching on a probe D3-brane sitting at an
intermediate position in the AdS bulk. It is shown that the
inclusion of angular momentum reduces the potential barrier thus
enhancing the Schwinger effect. Namely, producing particle pairs
would be easier in rotating medium, in accordance with previous
findings obtained from the local Lorentz transformation \cite{yz}.
Also, the results show that angular momentum has important effect
for the particle pair lying in the transversal plane compared with
that along the longitudinal orientation.

Moreover, the results may provide an estimate of how the Schwinger
effect changes with $\eta/s$ at strong coupling. From (\ref{eta})
one sees that $\eta_\perp/s$ is not affected by $a$ but
$\eta_\parallel/s$ decreases as $a$ increases. Here we will not
make much comment on why only one of the shear viscosities
saturates the bound, while the other may violate the bound (a
similar situation appeared in some anisotropic backgrounds
\cite{je,rc,ar}). We talk about $\eta_\parallel/s$. From the above
analysis one finds that increasing $a$ leads to decreasing
$\eta_\parallel/s$ thus making the fluid becomes more "perfect".
On the other hand, increasing $a$ leads to enhancing the Schwinger
effect. Taken together, one may conclude that at strong coupling
as $\eta/s$ decreases the Schwinger effect is enhanced.

However, there are some problems worthy for further studies.
First, here we just considered spinning Myers-Perry black holes
($a=b$), what will happen for general situation ($a\neq b$)?
Moreover, the potential analysis for Schwinger effect considered
here is basically within the Coulomb branch associated with the
leading exponent corresponding to the on-shell action of the
instanton. One can research the full decay rate if possible.

\section{Acknowledgments}
This work is supported by the Fundamental Research Funds for the
Central Universities, China University of Geosciences (Wuhan) with
No. G1323523064.


\end{document}